# General functions for human survival and mortality

Byung Mook Weon[1]

**Abstract**
General functions for human survival and mortality may support a possibility of general mechanisms in human ageing. We discovered that the survival and mortality curves could be described very simply and accurately by the Weibull survival function with age-dependent shape parameter. The age-dependence of shape parameter determines the shape of the survival and mortality curves and tells the nature of the ageing rate. Especially, the progression of shape parameter with age may be explained by the increase of interaction among vital processes or the evolution of susceptibility to faults with age. Age-related diseases may be attributed to the evolution of susceptibility to faults with age.

Survival function (showing the number of survivors as a function of age) is essential to understand the mechanisms of human ageing (*1,2,3*). Mortality function (indicating the relative rate for survival function decline) can be calculated by the derivative of survival function (*2,3*). Typical survival curves for humans show (i) a rapid decline in survival in the first few years of life and (ii) a relatively steady decline thereafter. Interestingly, the former curve shape resembles the Weibull survival function with the shape parameter smaller than unity and the latter curve shape seems to take after the case of the shape parameter larger than unity. With this in mind, we could assume that the shape parameter should be dependent upon age.

As stated above, we put forward for the first time general functions for survival (*S*) and mortality (*μ*) with age *t* as follows:

Survival: $\qquad S = \exp(-(t/\alpha)^{\beta(t)})$ (1)

Mortality: $\qquad \mu = \dfrac{d}{dt}((t/\alpha)^{\beta(t)})$ (2)

where *α* denotes the scale parameter and *β*(*t*) denotes the age-dependent shape parameter. It is noteworthy that the scale parameter (as known as the characteristic life) is always found at the characteristic survival or 'exp(-1)' ($\approx$36.79%), which therefore can be obtained *graphically* from the survival curve. After determining the value of *α*, since *β*(*t*) is mathematically equivalent to 'ln(-ln*S*)/ln(*t*/*α*)', from the plot of 'ln(-ln*S*)/ln(*t*/*α*)' with age, an adequate mathematical expression for *β*(*t*) can be obtained by regression analysis. If *β*(*t*) is constant with age, then the survival function is equal to the traditional Weibull survival function. After determining an adequate function for *β*(*t*), the mortality function can be calculated by the following equation:

Mortality: $\qquad \mu = (t/\alpha)^{\beta(t)} \times \left[ \dfrac{\beta(t)}{t} + \ln(t/\alpha) \times \dfrac{d\beta(t)}{dt} \right]$ (3)

Informatively, the age-dependence of shape parameter determines the shape of survival and mortality curves. This mortality function becomes the Weibull mortality function (suggested in 1951) for constant *β* and it seems to follow the Gompertz mortality law (suggested in 1825) at advanced age for variable *β*(*t*) (later shown in Fig. 1C).

The survival by age for the period 1900-1902 (open square) and 2000 (open circle) for all races (white, black, male, and female) in the United States (*4*) was evaluated in Fig. 1. Especially, we observed that when *α*=84.47, a linear relation for *β*(*t*) was appropriate for the case of 2000 (Fig. 1A). This linear function produced an accurate survival function with extremely high correlation (correlation coefficient, $r$=0.99998) between the actual data (open circle) and the estimated curve (solid line) (Fig. 1B). On the other hand, for the case of 1900-1902, when $α$=67.44, a quadratic relation for $β(t)$ was appropriate (Fig. 1A). There was also a very high correlation ($r$=0.99902) between the actual data (open square) and the estimated curve (solid line) (Fig. 1B).

A linear or quadratic expression for $β(t)$ is enough in this study. Such polynomial expression for $β(t)$ is convenient for the calculation of mortality by the equation (3). The adequate mathematical expression for $β(t)$ should be evaluated by the actual demographic data. The calculated mortality patterns for 1900-1902 and 2000 are illustrated in Fig. 1C. These patterns are consistent with the mortality patterns of other studies (*5,6,7*). Interestingly, the mortality profiles seem to follow the Gompertz law (the mortality increases *exponentially* with age) for late-life period (Fig. 1C, dashed line).

We should note that the quadratic term of $β(t)$ for 1900-1902 indicates a high juvenile mortality, corresponding to a low slope '**b**' of $β(t)$. Undoubtedly, the low juvenile mortality is important for longevity (*2,3*). To extend healthy lifespan, we should look towards higher values of $β(t)$ as the case of '**a>b**' in Fig. 1A. The age-dependence of shape parameter tells well the nature of the ageing rate.

The age-dependence of shape parameter must be associated with the fundamental mechanisms of human ageing. Especially, the shape parameter increases linearly with age for modern survival curve (for 2000). It is likely to imply a possibility that a general mechanism governs the human ageing through life. It is difficult that one can be aware of this fact from the traditional mortality studies. Furthermore, the extension of human lifespan maybe encounters a limit that at least is unsolved so far. For the case of the United States, since 1940's, the lifespan has increased by virtue of the continuous increase of the scale parameter (Fig. 2A), but the slope of $β(t)$ (for linear relation) has been almost invariant (Fig. 2B). The constant slope of $β(t)$ may be associated with the fundamental limits of longevity.

Ageing is a complex phenomenon (*3*). For example, it was reported that many genes act in concert to control longevity and ageing in complex biochemistry (*8*). Similarly, biological or physical processes sustaining life are extremely complex. It is of great importance that in the field of physics on complex systems, the decay phenomenon associated with multi-components often follows 'stretched exponential function' or 'Kohlrausch-Williams-Watts function' (*9,10*). Surprisingly, the mathematical expression for the stretched exponential function is identical with the Weibull survival function. In the stretched exponential function, the exponent is known as 'stretched exponent'. It is believed that the stretched exponent is associated with the simultaneous interactions with numerous elements or elemental processes (*10*). The progression of shape parameter may be explained by an assumption of that interaction becomes stronger with age. Variation of the nature of interaction among vital processes may be important to understand the nature of ageing process. If there is a stronger interaction at advanced age, then the body maybe becomes more susceptible to faults. In other words, the susceptibility to faults increases with age as the interaction among vital processes becomes strong with age. It is understandable for example that the recovery of a broken bone is easier for a child than for an adult. In this case, the susceptibility to faults is lower for the child than for the adult. Interestingly, for humans, the shape parameter is progressive with age, whereas for technical devices, the shape parameter is in general constant with age (time). Probably, humans (biological systems) are different from technical devices (artificial systems) in the nature of evolution of susceptibility with age. Age-related diseases may be attributed to the evolution of susceptibility to faults with age.

In summary, human survival and mortality curves could be described very simply and accurately by the Weibull survival function with the age-dependent shape parameter. The age-dependence of shape parameter determines the shape of survival and mortality curves and tells the nature of the ageing rate. Especially, the progression of shape parameter with age may be explained by the increase of interaction among vital processes or the evolution of susceptibility to faults with age. Age-related diseases may be attributed to the evolution of susceptibility to faults with age.

# References

1. Kirkwood T. B. L., Austad, S. N., 2000. Why do we age? Nature 408, 233-238.
2. Gavrilov, L. A., Gavrilova, N. S., 2003. Early-life factors modulating lifespan. In: Rattan, S. I. S. (Eds.) Modulating ageing and longevity. Kluwer Academic Publishers, Dordrecht, pp. 27-50.
3. Gavrilov, L. A., Gavrilova, N. S., 2003. The quest for a general theory of aging and longevity. Science's SAGE KE.
(http://sageke.sciencemag.org/cgi/content/full/sageke;2003/28/re5)
4. Arias, E., 2002. United States life tables, 2000. National Vital Statistics Reports. National Center for Health Statistics, Huattsville, Maryland. Vol. 51, No. 3.
5. Westendorp, R. G. J., Kirkwood, T. B. L., 1998. Human longevity at the cost of reproductive success. Nature 396, 743-746.
6. Azbel, M. Y., 1999. Phenomenological theory of mortality evolution: its singularities, universality, and superuniversality. Proc. Natl. Acad. Sci. U.S.A. 96, 3303-3307.
7. Lee, R. D. P., 2003. Rethinking the evolutionary theory of aging: transfer, not births, shape senescence in social species. Proc. Natl. Acad. Sci. U.S.A. 100, 9637-9642.
8. Gems, D., McElwee, J. J., 2003. Ageing: Microarraying mortality. Nature 424, 259-261.
9. Hunt, E. R., 2002. Stretched exponential dynamics in a chain of coupled chaotic oscillators. Gade, P. M., and Mousseau, N. Europhys. Lett. 60, 827-835.
10. Debenedetti, P. G., Stillinger, F. H., 2001. Supercooled liquids and the glass transition. Nature 410, 259-367.
11. Nelson, W., 1990. Accelerated testing: statistical functions, test plans, and data analyses. John Wiley & Sons, New York, pp. 63-65.

**Figure legends**

**FIG. 1.** Shape parameter (A), survival (B), and mortality (C) as a function of age. The survival by age for the period 1900-1902 (open square) and 2000 (open circle) for all races in the United States (*4*) was evaluated. From the age-dependent shape parameter, *β(t)*, the survival and mortality curves (solid line) were plotted by the general functions: $S = \exp(-(t/\alpha)^{\beta(t)})$ and $\mu = d[(t/\alpha)^{\beta(t)}]/dt$, respectively. Especially, for the most duration of life, a linear relation, '*β*(*t*)=1.43899+0.07280×*t*', for 2000 and a quadratic relation, '*β*(*t*)=0.56539+0.01558×*t*+0.0003348×*t*$^2$', for 1900-1902 were observed.

**FIG. 2.** Variations of the scale parameter (A) and the slope of shape parameter (B) (for linear relation) for the United States from 1900's to 2000's. Although the scale parameter has increased continuously, the slope of shape parameter has been almost invariant since 1940's.

**Figures**

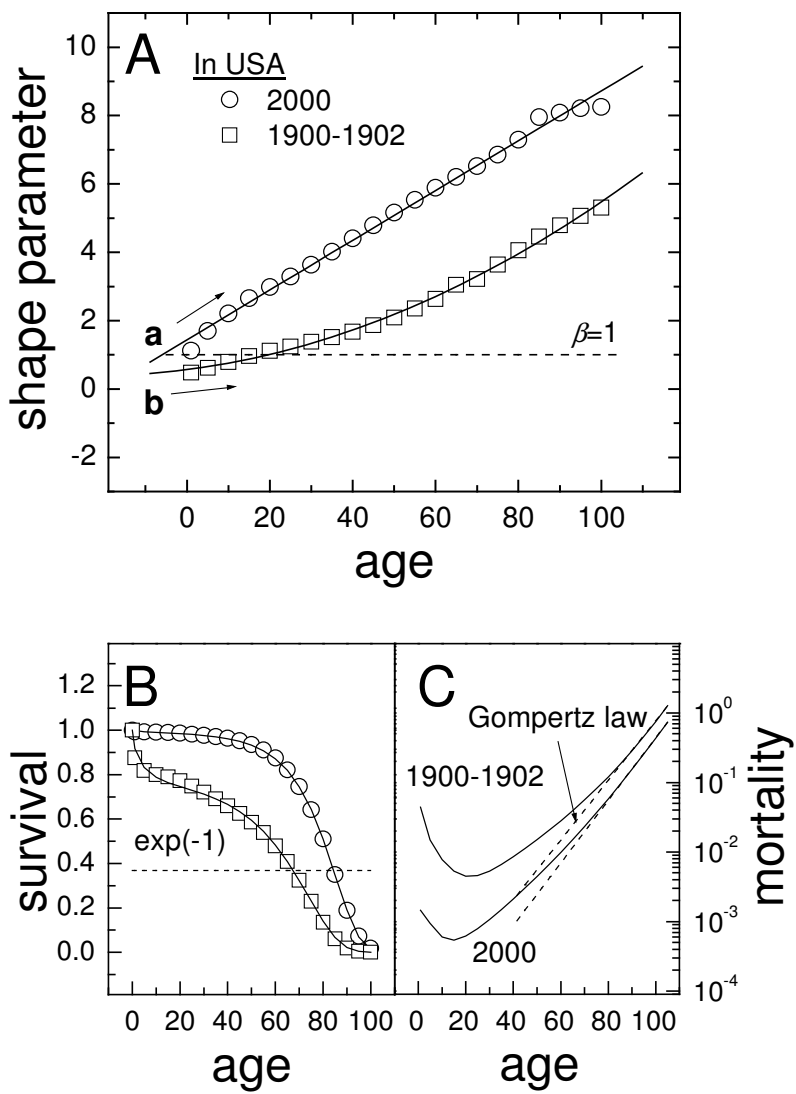

FIG. 1. B. M. Weon

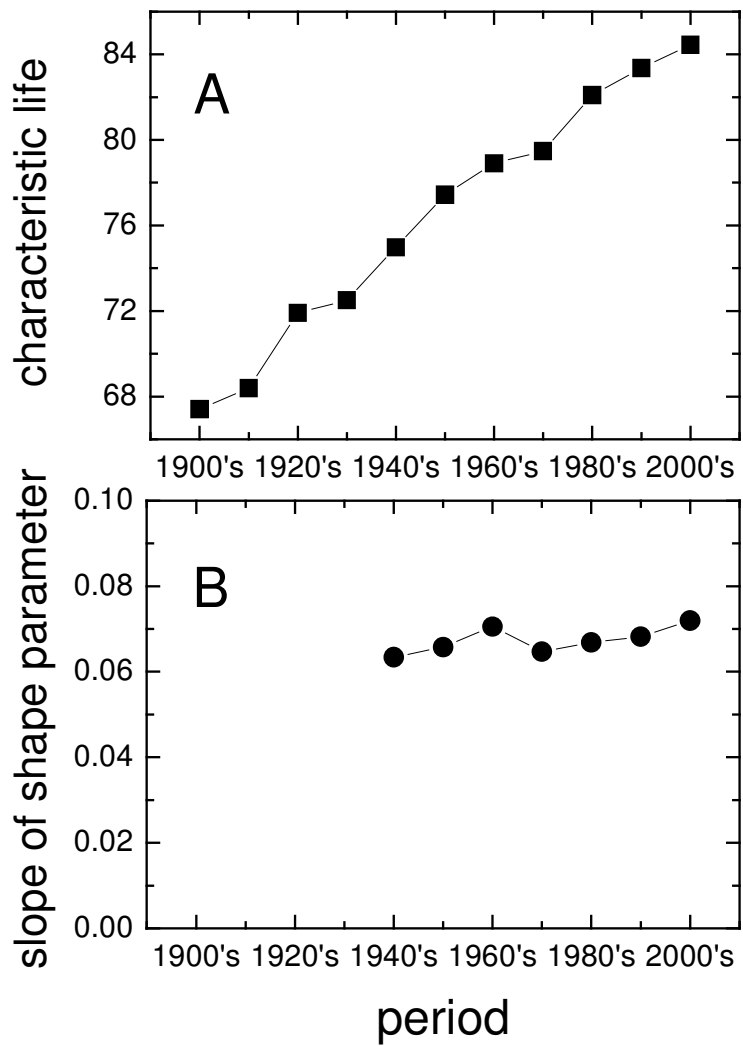

FIG. 2. B. M. Weon